\documentclass[russian,american]{article}
\usepackage[T1]{fontenc}
\usepackage{array}
\usepackage{float}
\usepackage{slashed}
\usepackage{multirow}
\usepackage{amsthm}
\usepackage{amsmath}
\usepackage{amssymb}
\usepackage{graphicx}

\makeatletter

\providecommand{\tabularnewline}{\\}
\floatstyle{ruled}
\newfloat{algorithm}{tbp}{loa}
\providecommand{\algorithmname}{Algorithm}
\floatname{algorithm}{\protect\algorithmname}

\newcommand{\lyxrightaddress}[1]{
\par {\raggedleft \begin{tabular}{l}\ignorespaces
#1
\end{tabular}
\vspace{1.4em}
\par}
}
  \theoremstyle{definition}
  \newtheorem{defn}{\protect\definitionname}
  \theoremstyle{plain}
  \newtheorem{cor}{\protect\corollaryname}
  \theoremstyle{plain}
  \newtheorem{lem}{\protect\lemmaname}
  \theoremstyle{remark}
  \newtheorem{rem}{\protect\remarkname}
\theoremstyle{plain}
\newtheorem{thm}{\protect\theoremname}
  \theoremstyle{plain}
  \newtheorem{prop}{\protect\propositionname}

\makeatother

\usepackage{babel}
  \addto\captionsamerican{\renewcommand{\definitionname}{Definition}}
  \addto\captionsamerican{\renewcommand{\lemmaname}{Lemma}}
  \addto\captionsamerican{\renewcommand{\propositionname}{Proposition}}
  \addto\captionsamerican{\renewcommand{\remarkname}{Remark}}
  \addto\captionsrussian{\renewcommand{\definitionname}{\inputencoding{koi8-r}Определение}}
  \addto\captionsrussian{\renewcommand{\lemmaname}{\inputencoding{koi8-r}Лемма}}
  \addto\captionsrussian{\renewcommand{\propositionname}{\inputencoding{koi8-r}Предложение}}
  \addto\captionsrussian{\renewcommand{\remarkname}{\inputencoding{koi8-r}Замечание}}
  \providecommand{\definitionname}{Definition}
  \providecommand{\lemmaname}{Lemma}
  \providecommand{\propositionname}{Proposition}
  \providecommand{\remarkname}{Remark}
\addto\captionsamerican{\renewcommand{\corollaryname}{Corollary}}
\addto\captionsamerican{\renewcommand{\theoremname}{Theorem}}
\addto\captionsrussian{\renewcommand{\algorithmname}{\inputencoding{koi8-r}Алгоритм}}
\addto\captionsrussian{\renewcommand{\corollaryname}{\inputencoding{koi8-r}Вывод}}
\addto\captionsrussian{\renewcommand{\theoremname}{\inputencoding{koi8-r}Теорема}}
\providecommand{\corollaryname}{Corollary}
\providecommand{\theoremname}{Theorem}

\begin{document}

\title{Polynomial Time Algorithm for Graph Isomorphism Testing}

\author{Michael I. Trofimov}

\maketitle

\lyxrightaddress{email: mt2@comtv.ru}
\begin{abstract}
\textit{This article deals with new polynomial time algorithm for
graph isomorphism testing.}

\end{abstract}

\section*{Introduction}

According to Harary's definition, two graphs are isomorphic if there
exists a one-to-one correspondence between their vertex sets which
preserves adjacency. In other words in language of matrix algebra,
two graphs with adjacency matrices $A$ and $A^{\prime}$ are isomorphic
iff there exists a permutation matrix $P$ such that $A=P^{-1}\cdot A^{\prime}\cdot P$.
\cite{Har}

In spite of efforts of many researchers, whether the problem of graph
isomorphism testing is NP-complete, is still open. There is an interesting
explanation of this fact in the book \cite{Garey}. The authors noted
that proofs of NP-completeness seem to require a certain amount of
redundancy, a redundancy that graph isomorphism problem lacks. For
example, in the case of subgraph isomorphism search the same result
may be observed even if some edges of given graph will be deleted
or some new edges will be added. In contrast, if any edge will be
added to one of two isomorphic graphs (or if any edge will be deleted),
then the graphs will no longer be isomorphic. So the graph isomorphism
problem is not typical NP-complete problem. 

Also it's important to note that number of effective polynomial algorithms
to test isomorphism of distinct graph classes were introduced. Particularly
such algorithm was implemented and proved for trees \cite{Aho,Hopcroft,Zykov}.

\section*{The general principles of the approach}

Further, without loss of generality of the task, we will consider
the solution for undirected connected graphs without loops \cite{Kobler,Miller}:

\[
G_{s}=(V,E_{s}),
\]

where $V$ is vertex set, $|V|=n$;

$E_{s}$ is edge set, $|E_{s}|=m$.

The graphs $G_{s}$ are called \textit{source graphs} or \textit{S-graphs}.

To avoid terminological misinterpretation let us recall some well-known
definitions that will be necessary further. \textit{Union of graphs}
$G\cup G^{\prime}$ is a graph that has vertex set $V\cup V^{\prime}$
and edge set $E\cup E^{\prime}$ \cite{Har}. Similarly, \textit{intersection
of graphs} $G\cap G^{\prime}$ is a graph that has vertex set $V\cap V^{\prime}$
and edge set $E\cap E^{\prime}$. Note that this definition assumes
''\textit{empty graph}'' with $V=\textrm{ь}$, $E=\textrm{ь}$. 

According to Harary's definition: ''Two points {[}vertices -- MT{]}
\textit{\emph{$u$ }}and \textit{\emph{$v$ }}of the graph \textit{\emph{$G$
}}are \textit{similar} if for some automorphism \textit{\emph{$\alpha$
}}of \textit{\emph{$G$, $\alpha(u)=v$}}'' \cite{Har}. We expand
this definition:
\begin{defn}
\textbf{\label{D1-similar-vert} }Vertex $v\in V$ of a graph $G=(V,E)$
and vertex $v^{\prime}\in V^{\prime}$ of a graph $G^{\prime}=(V^{\prime},E^{\prime})$
are \textit{similar, }if for some isomorphism $\pi$ of \textit{\emph{$G$
onto }}$G^{\prime}$, $\pi(v)=v^{\prime}$.\hfill{}$\square$
\end{defn}
Similar definition is possible for edges as well:
\begin{defn}
\textbf{\label{D2-similar-edges} }Edge $e$ of a graph $G$ and edge
$e^{\prime}$of a graph $G^{\prime}$ are \textit{similar, }if for
some isomorphism $\pi$ of \textit{\emph{$G$ onto }}$G^{\prime}$,
$\pi(e)=e^{\prime}$.\hfill{}$\square$

Also the following was noted: ''if $\alpha$ is an automorphism of
$G$, then it is clear that $G-u$ and $G-\alpha(u)$ are isomorphic.
Therefore if $u$ and $v$ are similar, then $G-u\cong G-v$ '' \cite{Har}.
Taking into account extended Definitions \ref{D1-similar-vert}, \ref{D2-similar-edges}
we have:\end{defn}
\begin{cor}
\label{=0004211-similar-vert}If $G\cong G'$ and vertices $v$ and
$v'$ are similar, then $G-v\cong G'-v'$.\hfill{}$\square$ 
\end{cor}
\phantom{}
\begin{cor}
\label{C2-similar-edges}If $G\cong G'$ and $\pi$ is some isomorphism
such that $e=\pi(e^{\prime}),\: e\in E,\: e^{\prime}\in E^{\prime}$,
then $G-e\cong G'-e'$ and $\pi$ is isomorphism for these graphs
also.\hfill{}$\square$ \end{cor}
\begin{lem}
\textbf{\label{L10-similar-adj-vertex-exists}}If $G\cong G^{\prime}$,
vertices $v\in V$ and $v^{\prime}\in V^{\prime}$ are similar, an
edge $(v,w)\in E$, then there is vertex $w^{\prime}\in V^{\prime}$
such that $w^{\prime}$ and $v^{\prime}$ are adjacent; edges $(v^{\prime},w^{\prime})\in E^{\prime}$
and $(v,w)$, vertices $w$ and $w^{\prime}$ are similar respectively.\hfill{}$\square$\end{lem}
\begin{proof}
Let graphs $G$ and $G^{\prime}$ be isomorphic, let vertices $v\in V$
and $v^{\prime}\in V^{\prime}$ be similar for some isomorphism $\pi:$
$v^{\prime}=\pi(v)$. Let $w_{1},w_{2},...,w_{k}\in V$ be adjacent
to $v$. Every other vertex of $G$ is not adjacent to $v$. Since
vertices $v$ and  $v^{\prime}$ are similar, we see that these vertices
have the same degree \cite{Chartrand}. Let $w_{1}^{\prime},w_{2}^{\prime},...,w_{k}^{\prime}\in V^{\prime}$
be adjacent to $v^{\prime}$. Every other vertex of $G^{\prime}$
is not adjacent to $v^{\prime}$. Suppose that there is vertex $w_{i},\:1\leqslant i\leqslant k$
for one is impossible to find similar vertex $w_{j}^{\prime},\:1\leqslant j\leqslant k$.
Then $x^{\prime}=\pi(w_{i}),\: x^{\prime}\in V^{\prime},\: x^{\prime}\notin\{w_{1}^{\prime},w_{2}^{\prime},...,w_{k}^{\prime}\}$.
Hence $x^{\prime}$ is not adjacent to $v^{\prime}$, but $w_{i}$
is adjacent to $v$, this means that mapping $\pi$ does not preserve
adjacency, thus $\pi$ is not isomorphism. This contradiction shows
that our supposition is not correct. Hence for vertex $w_{i}$ we
can always find similar vertex $w_{j}^{\prime}$, which vertex is
adjacent to $v^{\prime}$.
\end{proof}
Denote by $\mathrm{dist}(v,u)$ the distance between vertices $v,u\in V$.
If $v$ and $u$ are adjacent, then we say that $\mathrm{dist}(v,u)=1$.
Next corollary follows from Lemma \ref{L10-similar-adj-vertex-exists}: 
\begin{cor}
\label{C3-dist-for-similar-vert}If $G\cong G'$, vertices $v,w\in V$
and $v^{\prime},w^{\prime}\in V^{\prime}$ are similar respectively,
then $\mathrm{dist}(v,w)=\mathrm{dist}(v^{\prime},w^{\prime})$.\hfill{}$\square$ \end{cor}
\begin{lem}
\label{NL2}Let $G\cong G'$ and let $\pi$ be possible isomorphism
such that for non-adjacent vertices $v,w\in V$, $v^{\prime},w^{\prime}\in V^{\prime}$:
$v=\pi(v^{\prime})$, $w=\pi(w^{\prime})$. If we add edge $(v,w)$
to $G$ and add $(v^{\prime},w^{\prime})$ to $G^{\prime}$, then
we obtain isomorphic graphs $G_{2}$ and $G_{2}^{\prime}$, $\pi$
is possible isomorphism.\hfill{}$\square$\end{lem}
\begin{proof}
Since $\pi:\: G\cong G^{\prime}$, we have $\pi:\:\bar{G}\cong\bar{G}^{\prime}$,
where $\bar{G}=(V,\bar{E})$, $\bar{G}^{\prime}=(V^{\prime},\bar{E^{\prime}})$
are complementary graphs. Since the vertices $v,w$ (and $v^{\prime},w^{\prime}$)
are not adjacent within $G$ and $G^{\prime}$, we have edges $(v,w)\in\bar{E}$
and $(v^{\prime},w^{\prime})\in\bar{E^{\prime}}$. From Corollary
\ref{C2-similar-edges} it follows that $\pi:\:\bar{G}-(v,w)\cong\bar{G}'-(v',w^{\prime})$.
Let $\bar{G}-(v,w)=H$, $\bar{G^{\prime}}-(v^{\prime},w^{\prime})=H^{\prime}$;
then $\pi:\: H\cong H'$. Thus $\pi:\:\bar{H}\cong\bar{H}'$. But
$\bar{H}=(V,E\cup(v,w))=G_{2}$, $\bar{H^{\prime}}=(V^{\prime},E^{\prime}\cup(v^{\prime},w^{\prime}))=G_{2}^{\prime}$,
hence $\pi:\: G_{2}\cong G_{2}^{\prime}$.
\end{proof}
Let us \textbf{define procedure} \textbf{$P_{B}$}. We replace every
edge $(i,j)$ of source graph with \textit{additional vertex} $k$
and edges $(i,k)$, $(k,j)$. Assign orange color to every additional
vertex. Assign black color to every vertex of source graph.\hfill{}$\square$

In the result of \textbf{$P_{B}$ }we get bipartite graph, where orange
vertices are the first part and black vertices are the second part:

\[
G_{B}=(U,V,E),
\]

where $U$ is subset of the first part vertices (additional orange
vertices), $|U|=m$; 

$V$ is subset of the second part vertices (black vertices of source
graph), $|V|=n$; 

$E$ is edge set, $|E|=2m$.

The graphs $G_{B}$ are called $B$-\textit{graphs}. Note that each
orange vertex has degree 2 by construction. Also, we see that

\[
G_{B}\cong G_{B}^{\prime}\Longleftrightarrow G_{s}\cong G_{s}^{\prime}.
\]

\begin{flushleft}
Let us select an arbitrary vertex $t\in V$, called \textit{start
vertex,} and place all vertices of the graph by levels such that a
vertex $v$ is placed on level $d$, if $\mathrm{dist}(v,t)=d$ (Fig.\ref{fig:F1-D-alfa}).
\begin{figure}[H]
\includegraphics[scale=0.4]{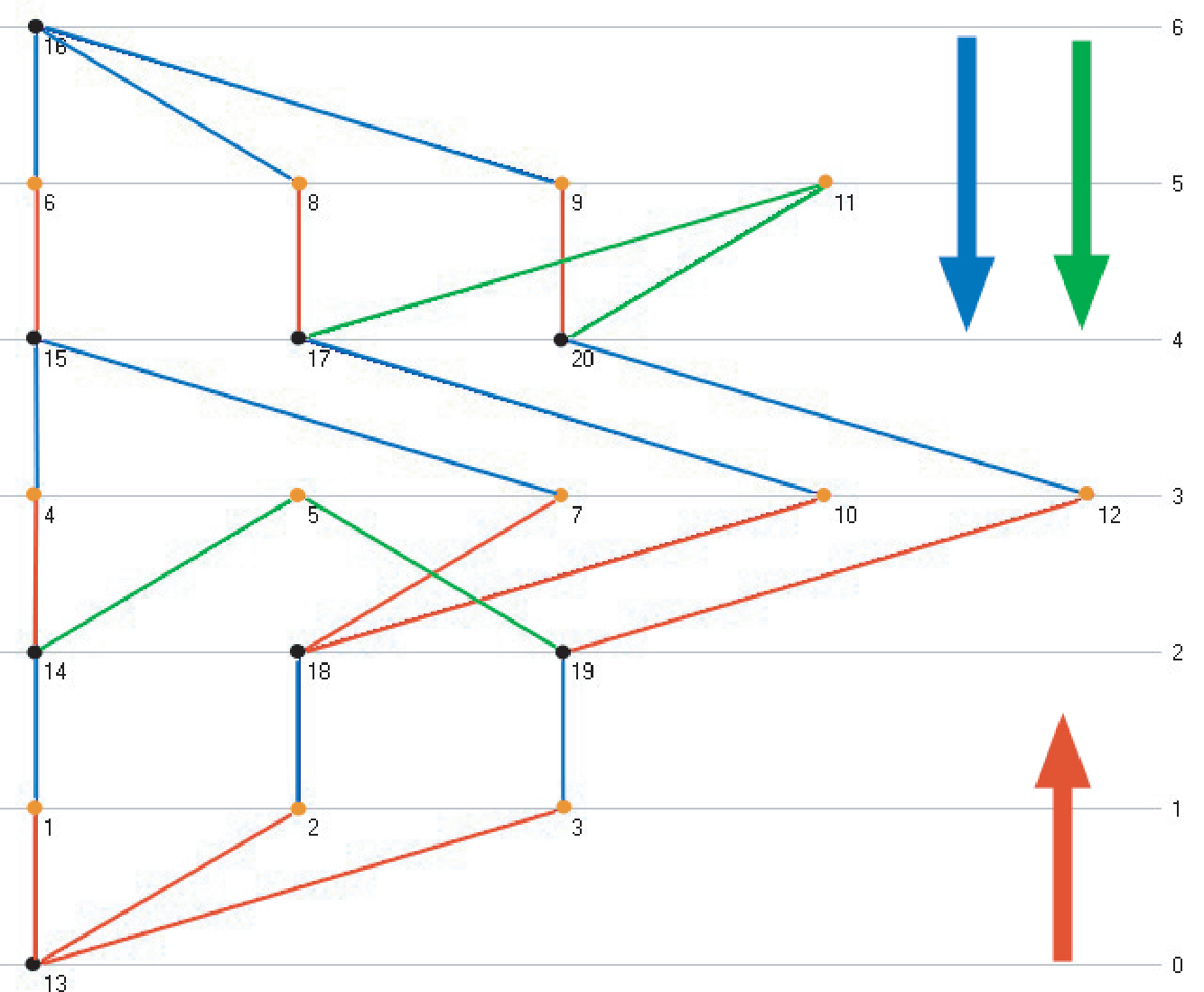}

\caption{$D_{\alpha}(t)$.\label{fig:F1-D-alfa}}
\end{figure}

\par\end{flushleft}

Note that we have only one vertex (i.e. start vertex $t$) on zero
level. We have black vertices from $V$ on levels $d=2k,\: k=0,1,2,..$
and we have orange vertices from $U$ on levels $d=2k+1$. Trees with
height 1 grow from black vertices of even level $d=2k$ (excluding
last level) to orange vertices of next level $d=2k+1$ (let us assign
red color to every edge of such tree). Also another one height trees
grow from black vertices of level $d=2k$ (excluding zero level) to
orange vertices of foregoing level $d=2k-1$ (let us assign blue color
to every edge of such tree). So every such tree has one black root
and one or more orange leafs. Also let us note trees which have two
black leafs on the same level and orange root on next level. Let us
assign green color to every edge of such tree. A red edge tree is
called \textit{red tree}. A blue edge tree is called \textit{blue
tree}. A green edge tree is called \textit{green tree}. All such (red,
blue, green) trees are called $\alpha$\textit{-trees. }The colored
arrows on Fig. \ref{fig:F1-D-alfa} show growing direction of $\alpha$-trees.
The procedure of $\alpha$-trees selection is called $\alpha$\textit{-decomposition}
of $B$-graph and denoted $D_{\alpha}(t)$. 
\begin{lem}
\label{L1-D-alfa}All edges of $B$-graph are colored in the result
of $D_{\alpha}$ .\hfill{}$\square$\end{lem}
\begin{proof}
Since $B$-graph is bipartite graph, we see that every edge ($i,j$)
is incident with one black vertex $i$ and one orange vertex $j$.
If $i$ is placed on a level higher than level of $j$, then $D_{\alpha}$
assigns blue color to the edge ($i,j$). If $i$ is placed on lower
level than level of $j$, then $D_{\alpha}$ assigns red or green
color to the edge ($i,j$). Another case is impossible.
\end{proof}
\begin{flushleft}
\begin{figure}[h]
\includegraphics[scale=0.4]{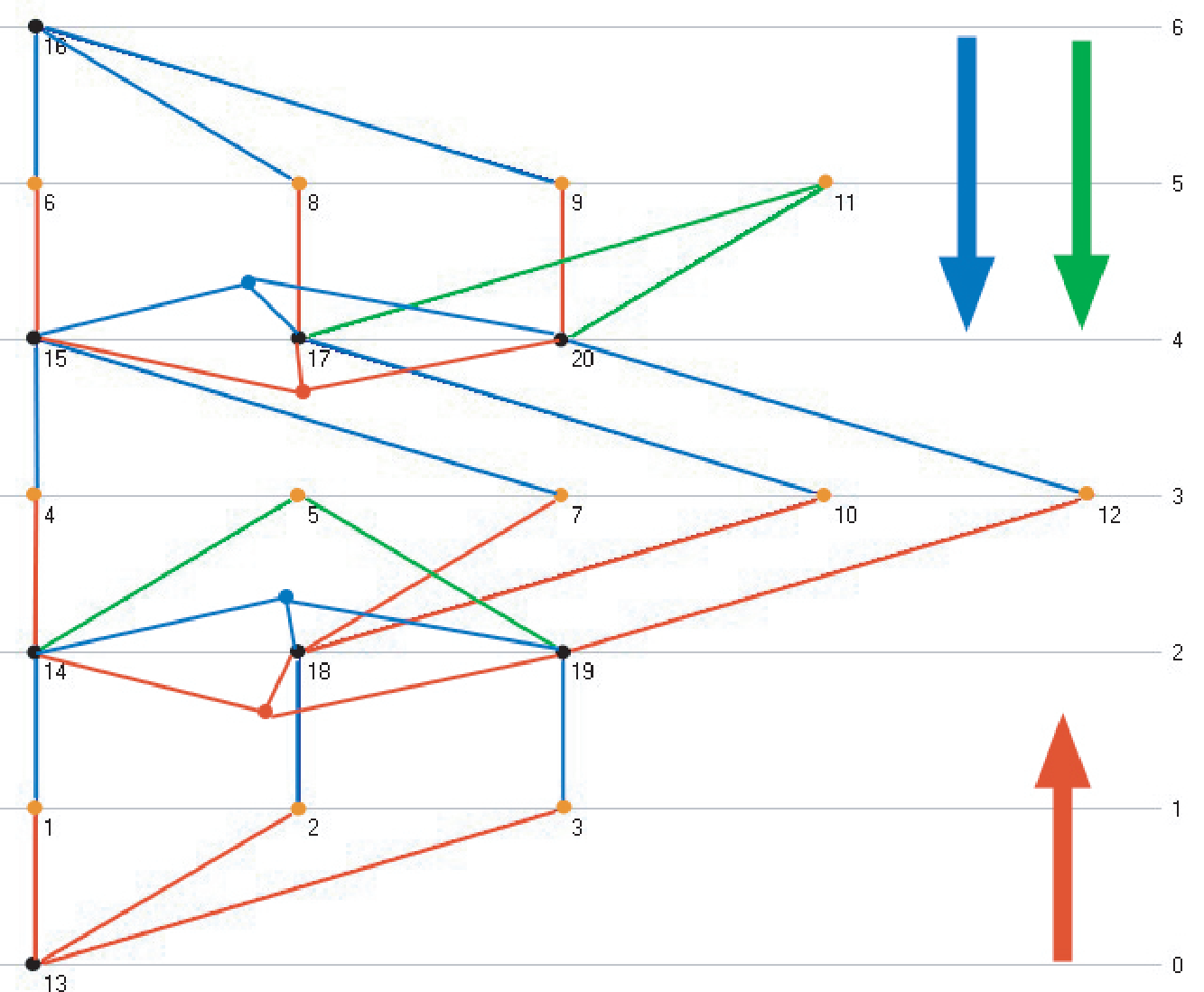}\caption{$D_{\beta}(t)$.\label{fig:F2-D-beta}}
\end{figure}

\par\end{flushleft}

If more than one of $\alpha$-trees of the same color grow on any
even level $d$, then we can unit these trees into one tree. For this
purpose we add \textit{additional vertex} and edges between this vertex
and roots of given trees. Let us assign color of the trees to additional
vertex (Fig. \ref{fig:F2-D-beta}). Produced tree is called $\beta$\textit{-tree}
of level $d$. If only one tree grows on any even level, then this
tree is called $\beta$-tree also. The procedure of $\beta$-trees
producing is called $\beta$\textit{-decomposition} of $B$-graph
and denoted $D_{\beta}(t)$, where $t$ is start vertex.
\begin{lem}
\label{L2-beta-tree}Let $B$-graphs $G_{B}=(U,V,E)$ and $G_{B}^{\prime}=(U^{\prime},V^{\prime},E^{\prime})$
be isomorphic and let vertices $t\in V,\: t^{\prime}\in V^{\prime}$
be similar. Let us produce $D_{\beta}(t)$ and $D_{\beta}(t^{\prime})$.
Then total number of levels $L$ of graph $G_{B}$ is the same as
the number of levels of graph $G_{B}^{\prime}$ and every $\beta$-tree
of level \textup{$d,\:0\leqslant d<L$} of graph $G_{B}$ and every
$\beta$-tree of level \textup{$d$} of graph $G_{B}^{\prime}$ with
the same color are isomorphic.\hfill{}$\square$\end{lem}
\begin{proof}
By construction we have not more than one $\beta$-tree with selected
color on every level $d$. Thus in the result of $D_{\beta}$ we get
a forest of red, blue and green trees. We see that tree distribution
by levels depends on start vertex choice only, i.e. the distribution
depends on distance to $t$, but does not depend on vertex numbering.
Suppose that any black vertex $v\in V$ of level $d_{1}$ and every
vertex of level $d_{1}$ of graph $G_{B}^{\prime}$ are not similar.
Since the graphs are isomorphic, we see that vertex $v$ and vertex
$v^{\prime}\in V^{\prime}$ of level $d_{2},\: d_{2}\neq d_{1}$ are
similar. But in this case, we obtain $\mathrm{dist}(t,v)\neq\mathrm{dist}(t^{\prime},v^{\prime})$,
this contradicts Corollary \ref{C3-dist-for-similar-vert}. Hence
our supposition is not correct and for every vertex $v\in V$ of level
$d_{i}$ we can find similar vertex $v^{\prime}\in V^{\prime}$ of
of level $d_{i}$. From Lemma \ref{L10-similar-adj-vertex-exists}
it follows that for every neighbor of $v$ we can find similar neighbor
of $v^{\prime}$, thus these $\beta$-trees are isomorphic.
\end{proof}
For data structures definition we will use the Backus -- Naur form
with standard metasymbols:

\[
<>::=\{\}|
\]
and with additional metasymbol:

\[
\diamondsuit
\]
this metasymbol means that sequence, followed behind it, is sorted
in descending order. We say about substrings sequence (or subsequences),
where every substring is considered as indissoluble instance. For
comparison of strings we will use following well-known rule: \textquotedblleft{}Given
two arrays $x$ and $y$ the relation $x<y$ holds if and only if
there exists an index $k$ such that $x[k]<y[k]$ and $x[i]=y[i]$
for all $i<k$'' \cite{Wirth}. If the arrays have different length,
then we have to add zeros (in the case of integer arrays) or space
symbols (in the case of strings) to the end of shorter array such
that the length of the arrays would be the same.
\begin{rem}
\label{N1-cmp-array}String comparison requires time proportional
to the length of the string.\hfill{}$\square$\\
\end{rem}
\begin{defn}
\label{D3-BNF}\-

\begin{tabular}{lc}
$<\mathrm{empty\: string}>::=$ & \tabularnewline
$<\mathrm{digit}>::=0|1|2|3|4|5|6|7|8|9$ & \tabularnewline
$<\mathrm{value}>::=<\mathrm{digit}>\{<\mathrm{digit}>\}$ & \tabularnewline
$<\mathrm{tuple}>::=<\mathrm{empty\: string}>|\diamondsuit<\mathrm{value}>\{,<\mathrm{value}>\}$ & (1)\tabularnewline
$<\mathrm{rgb\: label}>::=<\mathrm{tuple}>;<\mathrm{tuple}>;<\mathrm{tuple}>$ & (2)\tabularnewline
$<\mathrm{level}>::=<\mathrm{value}>$ & (3)\tabularnewline
$<\mathrm{degree}>::=<\mathrm{value}>$ & (4)\tabularnewline
$<\mathrm{simple\: vertex\: code}>::=<\mathrm{level}>;<\mathrm{degree}>;<\mathrm{rgb\: label}>$ & (5)\tabularnewline
$<\mathrm{vertex\: code}>::=<\mathrm{simple\: vertex\: code}>\diamondsuit\{*<\mathrm{edge\: code}>\}$ & (6)\tabularnewline
$<\mathrm{edge\: code}>::=(\diamondsuit<\mathrm{vertex\: code}>.<\mathrm{vertex\: code}>)$ & (7)\tabularnewline
$<\mathrm{vertex\: invariant}>::=\diamondsuit<\mathrm{vertex\: code}>\{\&<\mathrm{vertex\: code}>\}$ & (8)\tabularnewline
\end{tabular}

\hfill{}$\square$
\end{defn}
\medskip{}

The defined data structures support classical \textit{tuple technique}
for tree isomorphism testing \cite{Aho,Hopcroft,Zykov}. To use this
technique, we place tree vertices onto levels in dependence on their
distance from tree root. Moving from leafs to the root we set correspondence
between a vertex and a tuple: every leaf has the tuple which consists
one unit. A tuple of level $d$ has a form:

\[
p,\: s_{1},\: s_{2},\:...,\: s_{k},
\]
where $s_{1},\: s_{2},\:...,\: s_{k}$ are vertex tuple of foregoing
level, $s_{i}\geqslant s_{i+1},\: i=1,...,k-1;$ 

$p=1+\underset{j=1}{\overset{k}{\sum}}s_{j}[1];$

$s_{j}[1]$ is the first element of tuple $s_{j}$.

Note that this technique may be used for labeled tree also. In this
case, vertex label should to be added to the tuple \cite{Aho}. Taking
into account Remark \ref{N1-cmp-array}, we obtain following theorem: 
\begin{thm}
\label{T1-tuple-time}The tuple technique requires linear time proportional
to the number of given tree vertices \cite{Aho,Hopcroft}.\hfill{}$\square$ 
\end{thm}
Also it was proved that equality of central tuples is necessary and
sufficient condition for trees isomorphism \cite{Zykov}. Since we
do not consider bicenter trees, we can reform this theorem as following: 
\begin{thm}
\textbf{\label{T2-tree-iso}}Equality of root tuples is necessary
and sufficient condition for rooted trees isomorphism.\hfill{}$\square$\end{thm}
\begin{cor}
\label{CC4}Tuple of a vertex uniquely represents tree (subtree),
where this vertex is a root.\hfill{}$\square$
\end{cor}
It is important to note the difference between tree levels for tuple
technique usage (from the leafs to the root) and levels of $B$-graph,
where red and blue trees grow to counter-directions. It will be clear
from a context, what level type we mean.
\begin{lem}
\textbf{\label{L7-similar-tree-vert}} Vertices of two isomorphic
trees are similar iff their tuples and pairs of tuples of their ancestors
(up to the root) are equal respectively.\hfill{}$\square$\end{lem}
\begin{proof}
Let us consider trees $S_{0}=(V,E)$ and $S_{0}^{\prime}=(V^{\prime},E^{\prime})$
with the roots $r$ and $r^{\prime}$ respectively (Fig. \ref{fig:F5-eq-tuples}).

\begin{figure}[H]
\includegraphics[scale=0.4]{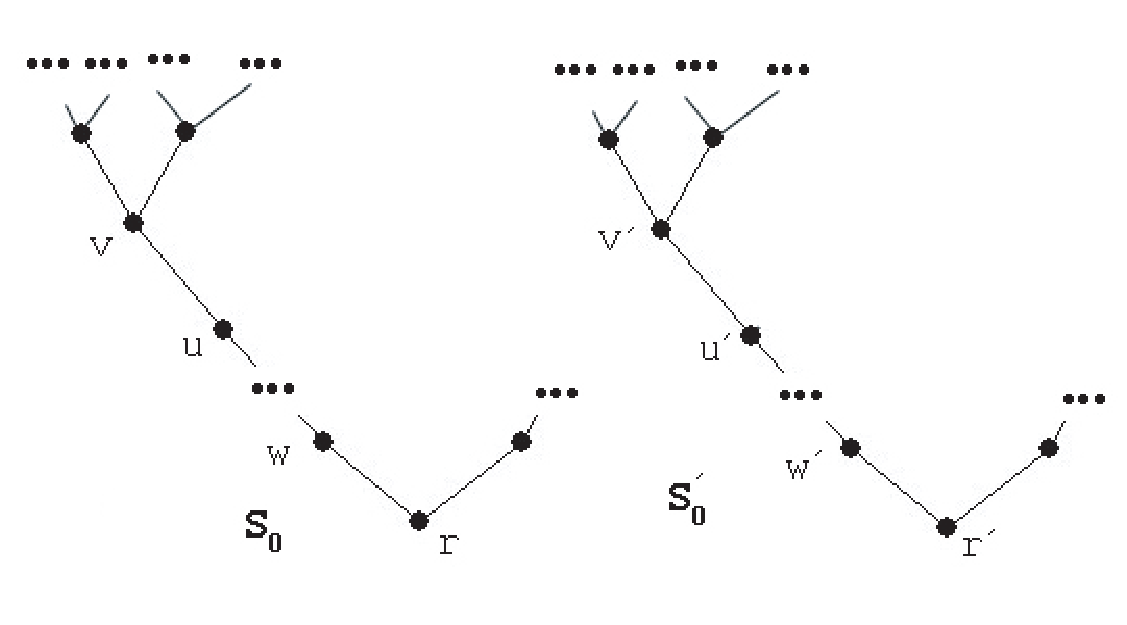}

\caption{Equal tuples of ancestors.\label{fig:F5-eq-tuples}}
\end{figure}

Let the tuples of vertices $v$ and $v^{\prime}$, and pairs of tuples
of their ancestors $u,...,w,r$ and $u^{\prime},...,w^{\prime},r^{\prime}$
are equal respectively. Since tuples of the roots $r$ and $r^{\prime}$
are equal, we obtain $S_{0}\cong S_{0}^{\prime}$ and $r\leftrightarrow r^{\prime}$
from Theorem \ref{T2-tree-iso}. Removing $r$ and $r^{\prime}$,
we obtain subtrees $S_{1}$ and $S_{1}^{\prime}$ with the roots $w$
and $w^{\prime}$. The root tuples are equal by condition, hence these
subtrees are isomorphic and $w\leftrightarrow w^{\prime}$. Continuing
root removing, after $k$-th removing, we get subtrees $S_{k}$ and
$S_{k}^{\prime}$ with the roots $v$ and $v^{\prime}$. Their tuples
are equal, hence these subtrees are isomorphic and $v\leftrightarrow v^{\prime}$.
Let us consider produced sequences of isomorphic trees $S_{k},..,S_{0}$
and $S_{k}^{\prime},...,S_{0}^{\prime}$ as separated graphs. The
root tuples of trees $S_{k-1}$ and $S_{k-1}^{\prime}$ may be represented
as

\[
s_{u}=(p+s_{v}[1],\: h_{1},...,\: h_{x},\: s_{v},\: t_{1},...,t_{y}),
\]

where $h_{1},...,\: h_{x}$ are tuples of foregoing level such that
$h_{i}>s_{v},\: i=1,...,x$, (may be absent);

$s_{v}$ is the tuple of vertex $v$;

$t_{1},...,t_{y}$ are tuples of foregoing level such that $t_{i}\leqslant s_{v},\: i=1,...,y$,
(may be absent);

$p=1+\underset{j=1}{\overset{x}{\sum}}h_{j}[1]+\underset{j=1}{\overset{y}{\sum}}t_{j}[1];$

$h_{j}[1]$ is the first element of tuple $h_{j}$;

$t_{j}[1]$ is the first element of tuple $t_{j}$;

$s_{v}[1]$ is the first element of tuple $s_{v}$.

Similarly, the root tuple of tree $S_{k-1}^{\prime}$ may be represented
as

\[
s_{u}^{\prime}=(p^{\prime}+s_{v}^{\prime}[1],\: h_{1}^{\prime},...,\: h_{x}^{\prime},\: s_{v}^{\prime},\: t_{1}^{\prime},...,t_{y}^{\prime}).
\]

Removing edges $(v,u)$ and $(v^{\prime},u^{\prime})$ from trees
$S_{k-1}$ and $S_{k-1}^{\prime}$, we obtain two forests $F_{k-1}$
and $F_{k-1}^{\prime}$. There are two trees in every of these forests:
the tree $S_{k}$ (or $S_{k}^{\prime}$ for the second forest) and
the tree $Q_{k-1}$ (or $Q_{k-1}^{\prime}$) with the root $u$ (or
$u^{\prime}$). Taking into account $s_{v}=s_{v}^{\prime}$ and $s_{u}=s_{u}^{\prime}$,
we see that root tuples of trees $Q_{k-1}$ and $Q_{k-1}^{\prime}$
are equal. Hence, $Q_{k-1}\cong Q_{k-1}^{\prime}$, $F_{k-1}\cong F_{k-1}^{\prime}$
and an isomorphism with correspondences $v\leftrightarrow v^{\prime}$,
$u\leftrightarrow u^{\prime}$ is possible. Adding edges $(v,u)$
and $(v^{\prime},u^{\prime})$ to the trees $F_{k-1}$ and $F_{k-1}^{\prime}$,
we see that for isomorphism of the trees $S_{k-1}$ and $S_{k-1}^{\prime}$
the same correspondences $v\leftrightarrow v^{\prime}$, $u\leftrightarrow u^{\prime}$
are possible. Indeed, this follows from Lemma \ref{NL2}. Further,
let us repeat similar reasoning for trees $S_{k-2}$ and $S_{k-2}^{\prime}$
etc., until we come to trees $S_{0}$ and $S_{0}^{\prime}$, where
after removing edges $(w,r)$ and $(w^{\prime},r^{\prime})$, we obtain
two forests $F_{0}$ and $F_{0}^{\prime}$. There are two trees in
every of these forests: the tree $S_{1}$ (or $S_{1}^{\prime}$) and
the tree $Q_{0}$ (or $Q_{0}^{\prime}$) with the root $r$ (or $r^{\prime}$).
The root tuples of the trees $Q_{0}$ and $Q_{0}^{\prime}$ are equal.
Hence, $Q_{0}\cong Q_{0}^{\prime}$ and $F_{0}\cong F_{0}^{\prime}$
and isomorphism with correspondences $r\leftrightarrow r^{\prime}$,
$w\leftrightarrow w^{\prime}$, ...,$v\leftrightarrow v^{\prime}$
is possible. Adding edges $(w,r)$ and $(w^{\prime},r^{\prime})$
to the trees $F_{0}$ and $F_{0}^{\prime}$, we see that for isomorphism
of the trees $S_{0}$ and $S_{0}^{\prime}$ the same correspondences
are possible. Hence, vertices $v$ and $v^{\prime}$ are similar. 

Now, let the tuples of vertices $u$ and $u^{\prime}$ are not equal
(Fig. \ref{fig:F6-Ne-tuples}).

\begin{figure}[H]
\includegraphics[scale=0.4]{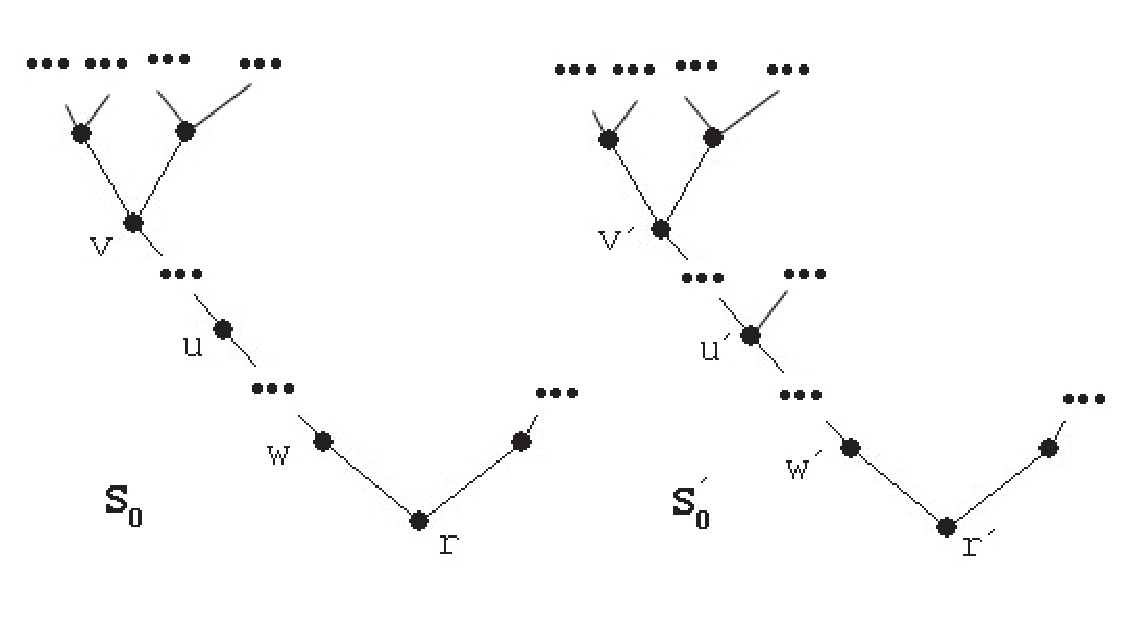}

\caption{Non equal tuples of ancestors.\label{fig:F6-Ne-tuples}}
\end{figure}

Since the tuples of the roots $r$ and $r^{\prime}$ are equal, we
see that $S_{0}\cong S_{0}^{\prime}$ and $r=\pi_{0}(r^{\prime})$,
where $\pi_{0}$ is isomorphism. Removing $r$ and $r^{\prime}$,
we obtain the trees with the roots $w$ and $w^{\prime}$. The root
tuples of these subtrees are equal. Hence, these subtrees are isomorphic
and $w=\pi_{1}(w^{\prime})$, where $\pi_{1}$is isomorphism. Continuing
such root removing, we come to subtrees with the roots $u$ and $u^{\prime}$.
The tuples of these roots are not equal. Hence these trees are not
isomorphic and vertices $v$ and $v^{\prime}$ are not similar.
\end{proof}
\begin{flushleft}
Let us consider graphs consist of two subgraphs $H_{1}$ and $H_{2}$.
These subgraphs have common vertices $v_{1},v_{2},...,v_{k}$ such
that an edge $(v_{i},v_{j}),\: i,j=1,2,...,k$ does not exist (Fig.
\ref{fig:F4-2}). Let us prove following lemma for such graphs. 
\begin{figure}[H]
\includegraphics[scale=0.4]{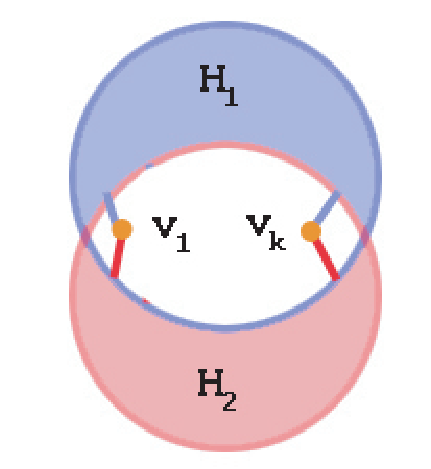}

\caption{Two subgraphs with common vertices within a graph. \label{fig:F4-2}}

\end{figure}

\par\end{flushleft}
\begin{lem}
\textbf{\label{L5-x-graphs}} If

1) $G=(V,E),$ $G^{\prime}=(V^{\prime},E^{\prime}),$

2) $G=H_{1}\cup H_{2}$, $G^{\prime}=H_{1}^{\prime}\cup H_{2}^{\prime}$,
$H_{1}=(V_{1},E_{1})$, $H_{2}=(V_{2},E_{2})$, $H_{1}^{\prime}=(V_{1}^{\prime},E_{1}^{\prime})$,
$H_{2}^{\prime}=(V_{2}^{\prime},E_{2}^{\prime})$,

3) $H_{1}\cap H_{2}=(X,\textrm{ь)}$, $X=\{v_{1},v_{2},...,v_{k}\},\: v_{i}\in V$,
$i=1,2,...,k$, $H_{1}^{\prime}\cap H_{2}^{\prime}=(X^{\prime},\slashed{O})$,
$H_{1}^{\prime}\cap H_{2}^{\prime}=\{v_{1}^{\prime},v_{2}^{\prime},...,v_{k}^{\prime}\}=X^{\prime},\: v_{i}^{\prime}\in V^{\prime}$,

4) Edges $(v_{i},v_{j}),\:(v_{i}^{\prime},v_{j}^{\prime}),\, v_{i},v_{j}\in X,\: v_{i}^{\prime},v_{j}^{\prime}\in X^{\prime},\: i,j=1,2,...,k$
do not exist,

5) $H_{1}\cong H_{1}^{\prime}$, $H_{2}\cong H_{2}^{\prime}$, and
$\Pi_{1},\:\Pi_{2}$ are isomorphism sets respectively,

6) $\exists\pi_{1},\:\pi_{2}:\:\pi_{1}\in\Pi_{1},\:\pi_{2}\in\Pi_{2}$
such that $v_{i}=\pi_{1}(v_{i}^{\prime})$, $v_{i}=\pi_{2}(v_{i}^{\prime})$,
$v_{i}\in X,\: v_{i}^{\prime}\in X^{\prime},\: i=1,2,...,k$,

then $G\cong G^{\prime}$.\hfill{}$\square$\end{lem}
\begin{proof}
Let $U_{1}=V_{1}\setminus X$, $U_{2}=V_{2}\setminus X$, $U_{1}^{\prime}=V_{1}^{\prime}\setminus X^{\prime}$,
$U_{2}^{\prime}=V_{2}^{\prime}\setminus X^{\prime}$, $p=|U_{1}|$,
$q=|U_{2}|$, $s=k+p+q$. Since $H_{1}\cong H_{1}^{\prime}$ by condition
5, we have $|U_{1}^{\prime}|=p$. Since $H_{2}\cong H_{2}^{\prime}$,
we have $|U_{2}^{\prime}|=q$. Without loss of generality, we can
assume that graphs vertices have following order of numbers: vertices
from $X$, $U_{1}$, $U_{2}$. Similarly for $G^{\prime}$: vertices
from $X^{\prime}$, $U_{1}^{\prime}$, $U_{2}^{\prime}$. Then $s\times s$
adjacency matrix of the first graph has following form:

\[
A=\left(\begin{array}{ccc}
O_{1} & B_{1} & B_{2}\\
B_{1}^{T} & C_{1} & O_{2}\\
B_{2}^{T} & O_{3} & C_{2}
\end{array}\right),
\]

\begin{flushleft}
where block $O_{1}$ is all-zero (by condition 4) $k\times k$ matrix; 
\par\end{flushleft}

\begin{flushleft}
block $B_{1}$ is $k\times p$ matrix corresponded to edges $(v,u),\: v\in X,\: u\in U_{1}$; 
\par\end{flushleft}

\begin{flushleft}
block $B_{2}$ is $k\times q$ matrix corresponded to edges $(v,u),\: v\in X,\: u\in U_{2}$; 
\par\end{flushleft}

\begin{flushleft}
block $C_{1}$ is $p\times p$ matrix corresponded to edges $(v,u),\: v,u\in U_{1}$; 
\par\end{flushleft}

\begin{flushleft}
block$C_{2}$ is $q\times q$ matrix corresponded to edges $(v,u),\: v,u\in U_{2}$; 
\par\end{flushleft}

\begin{flushleft}
blocks $O_{2}$ and $O_{3}$ are all-zero (since edge $(v,u)$, $v\in U_{1},\: u\in U_{2}$
does not exist by condition 3) $p\times q$ and $q\times p$ matrices
(Table \ref{tab:T1}).
\begin{table}[H]
\begin{tabular}{|l|c|c|c|c|}
\hline 
\multicolumn{1}{|l}{Vertex} & \multirow{2}{*}{} & \multirow{2}{*}{$X$} & \multirow{2}{*}{$U_{1}$} & \multirow{2}{*}{$U_{2}$}\tabularnewline
\multicolumn{1}{|l}{ subset} &  &  &  & \tabularnewline
\cline{2-5} 
 & Power & $k$ & $p$ & $q$\tabularnewline
\hline 
$X$ & $k$ & $\begin{array}{c}
O_{1}\\
k\times k
\end{array}$ & $\begin{array}{c}
B_{1}\\
k\times p
\end{array}$ & $\begin{array}{c}
B_{2}\\
k\times q
\end{array}$\tabularnewline
\hline 
$U_{1}$ & $p$ & $\begin{array}{c}
B_{1}^{T}\\
p\times k
\end{array}$ & $\begin{array}{c}
C_{1}\\
p\times p
\end{array}$ & $\begin{array}{c}
O_{2}\\
p\times q
\end{array}$\tabularnewline
\hline 
$U_{2}$ & $q$ & $\begin{array}{c}
B_{2}^{T}\\
q\times k
\end{array}$ & $\begin{array}{c}
O_{3}\\
q\times p
\end{array}$ & $\begin{array}{c}
C_{2}\\
q\times q
\end{array}$\tabularnewline
\hline 
\end{tabular}

\caption{\label{tab:T1}The blocks of matrix $A$ and their sizes.}
\end{table}

\par\end{flushleft}

Similar form has adjacency matrix of graph $G^{\prime}$:

\[
A^{\prime}=\left(\begin{array}{ccc}
O_{1} & B_{1}^{\prime} & B_{2}^{\prime}\\
B_{1}^{\prime T} & C_{1}^{\prime} & O_{2}\\
B_{2}^{\prime T} & O_{3} & C_{2}^{\prime}
\end{array}\right).
\]

To support vertex numbering agreement we can represent adjacency matrices
of subgraphs $H_{1}$ and $H_{2}$ in $s\times s$ matrix form. For
this purpose we can add all vertices from $U_{2}$ to set of vertices
of graph $H_{1}$. Similarly, we can add all vertices from $U_{1}$
to set of vertices of graph $H_{2}$. Clearly that such addition of
isolated vertices does not change adjacency. Thus, we obtain:

\[
A_{1}=\left(\begin{array}{ccc}
O_{1} & B_{1} & O\\
B_{1}^{T} & C_{1} & O_{2}\\
O & O_{3} & O
\end{array}\right),
\]

\[
A_{2}=\left(\begin{array}{ccc}
O_{1} & O & B_{2}\\
O & O & O_{2}\\
B_{2}^{T} & O_{3} & C_{2}
\end{array}\right),
\]

where $O$ are all-zero blocks of respective size.

Similar forms have adjacency matrices of subgraphs $H_{1}^{\prime}$
and $H_{2}^{\prime}$ after isolated vertices addition: 

\[
A_{1}^{\prime}=\left(\begin{array}{ccc}
O_{1} & B_{1}^{\prime} & O\\
B_{1}^{\prime T} & C_{1}^{\prime} & O_{2}\\
O & O_{3} & O
\end{array}\right),
\]

\[
A_{2}^{\prime}=\left(\begin{array}{ccc}
O_{1} & O & B_{2}^{\prime}\\
O & O & O_{2}\\
B_{2}^{\prime T} & O_{3} & C_{2}^{\prime}
\end{array}\right).
\]

As we can see, $A=A_{1}+A_{2}$ and $A^{\prime}=A_{1}^{\prime}+A_{2}^{\prime}$. 

Let $P$ be $s\times s$ permutation matrix for isomorphisms $\pi_{1},\:\pi_{2}$:
$A_{1}=P^{-1}\cdot A_{1}^{\prime}\cdot P$ and $A_{2}=P^{-1}\cdot A_{2}^{\prime}\cdot P$
. By condition 6 this matrix replaces rows and columns (corresponded
to common vertices $X^{\prime}$) of adjacency matrices of the subgraphs
to preserve adjacency. The result of the permutations for vertices
from $U_{2}^{\prime}$ in matrix $A_{1}^{\prime}$ is permutations
of zero rows and zero columns. The same result we obtain for permutations
for vertices from $U_{1}^{\prime}$ in matrix $A_{2}^{\prime}$. 

In the result we obtain:

\[
A=A_{1}+A_{2}=P^{-1}\cdot A_{1}^{\prime}\cdot P+P^{-1}\cdot A_{2}^{\prime}\cdot P=P^{-1}\cdot(A_{1}^{\prime}+A_{2}^{\prime})\cdot P=
\]
\[
P^{-1}\cdot\left(\begin{array}{ccc}
\left(\begin{array}{ccc}
O_{1} & B_{1}^{\prime} & O\\
B_{1}^{\prime T} & C_{1}^{\prime} & O_{2}\\
O & O_{3} & O
\end{array}\right) & + & \left(\begin{array}{ccc}
O_{1} & O & B_{2}^{\prime}\\
O & O & O_{2}\\
B_{2}^{\prime T} & O_{3} & C_{2}^{\prime}
\end{array}\right)\end{array}\right)\cdot P=
\]
\[
P^{-1}\cdot\left(\begin{array}{ccc}
0 & B_{1}^{\prime} & B_{2}^{\prime}\\
B_{1}^{\prime T} & C_{1}^{\prime} & 0\\
B_{2}^{\prime T} & 0 & C_{2}^{\prime}
\end{array}\right)\cdot P=P^{-1}\cdot A^{\prime}\cdot P.
\]

Hence, $G\cong G^{\prime}$.
\end{proof}
\hphantom{}

\begin{flushleft}
By $B$-graph construction every black vertex $v\in V$ may be root
of not more than three $\alpha$-trees of different colors. So to
characterize uniquely a vertex we write down tuples of red (r), blue
(b) and green (g) $\alpha$-trees with the root $v$ (Corollary \ref{CC4}).
In the result we obtain \textit{rgb} \textit{label} (2) (see Definition
\ref{D3-BNF}). Further we add \textit{level} $d$ (3) and vertex
\textit{degree} (4) to beginning of the string. In the result we obtain
\textit{simple vertex code} (5) for every black vertex. Now for every
edge of source graph $G_{s}$ we write down \textit{edge code} (7).
And again we produce \textit{vertex code} (6) for every vertex, but
now we take into account the codes of incident edges. From iteration
to iteration ''vertex code producing -- edge code producing'' vertex
code of every vertex reflects information about more and more remote
vertices and edges. To collect the information about all vertices
and all edges in vertex code we have to do $\mathrm{dm}(G)$ iterations,
where $\mathrm{dm}(G)$ is graph diameter. Getting every black vertex
as start vertex for decomposition we sort results. It produces \textit{vertex
invariants} (8) of the graph, this invariant is independent of vertex
numbering. This process is called procedure $P_{C}$ (Algorithm \ref{alg:A1}).
\begin{algorithm}[H]
\begin{enumerate}
\item call procedure $P_{B}$ for the source graph $G_{s}$, in the result
we obtain graph $G_{B}$; 
\item \textbf{for} $i:=1$ \textbf{to} $n$ \textbf{do}
\item \textbf{\quad{}begin}
\item \textbf{\quad{}\quad{}}call $D_{\alpha}(v_{i}),\: v_{i}\in V$; 
\item \textbf{\quad{}\quad{}}for every black vertex of graph $G_{B}$
produce $\mathrm{rgb}$ label (2) (see Definition \ref{D3-BNF}); 
\item \textbf{\quad{}\quad{}}for every black vertex of graph $G_{B}$
produce simple vertex code (5), assign this code to vertex code (6); 
\item \textbf{\quad{}\quad{}for} $j:=1$ \textbf{to} $\mathrm{dm}(G_{s})$\textbf{
do}
\item \textbf{\quad{}\quad{}\quad{}begin}
\item \textbf{\quad{}\quad{}\quad{}\quad{}}for every edge of graph $G_{s}$
produce edge code (7);
\item \textbf{\quad{}\quad{}\quad{}\quad{}}for every vertex of graph
$G_{s}$ produce vertex code (6) taking into account the codes of
incident edges; 
\item \textbf{\quad{}\quad{}\quad{}end;}
\item \textbf{\quad{}\quad{}}assign $C[j,i]:=c(v_{j})$, where $c(v_{j})$
is vertex code $v_{j}\in V$; 
\item \textbf{\quad{}end;}
\item sort every row of matrix $C$; 
\item sort matrix $C$ by rows;
\item \begin{raggedright}
for every $i$-th row of matrix $C$ produce vertex invariant (8),
assign it to $i$-th coordinate of vector $S$; 
\par\end{raggedright}
\end{enumerate}
\caption{\label{alg:A1}Procedure $P_{C}$.}
\end{algorithm}

\par\end{flushleft}

\begin{figure}[H]
\includegraphics[scale=0.4]{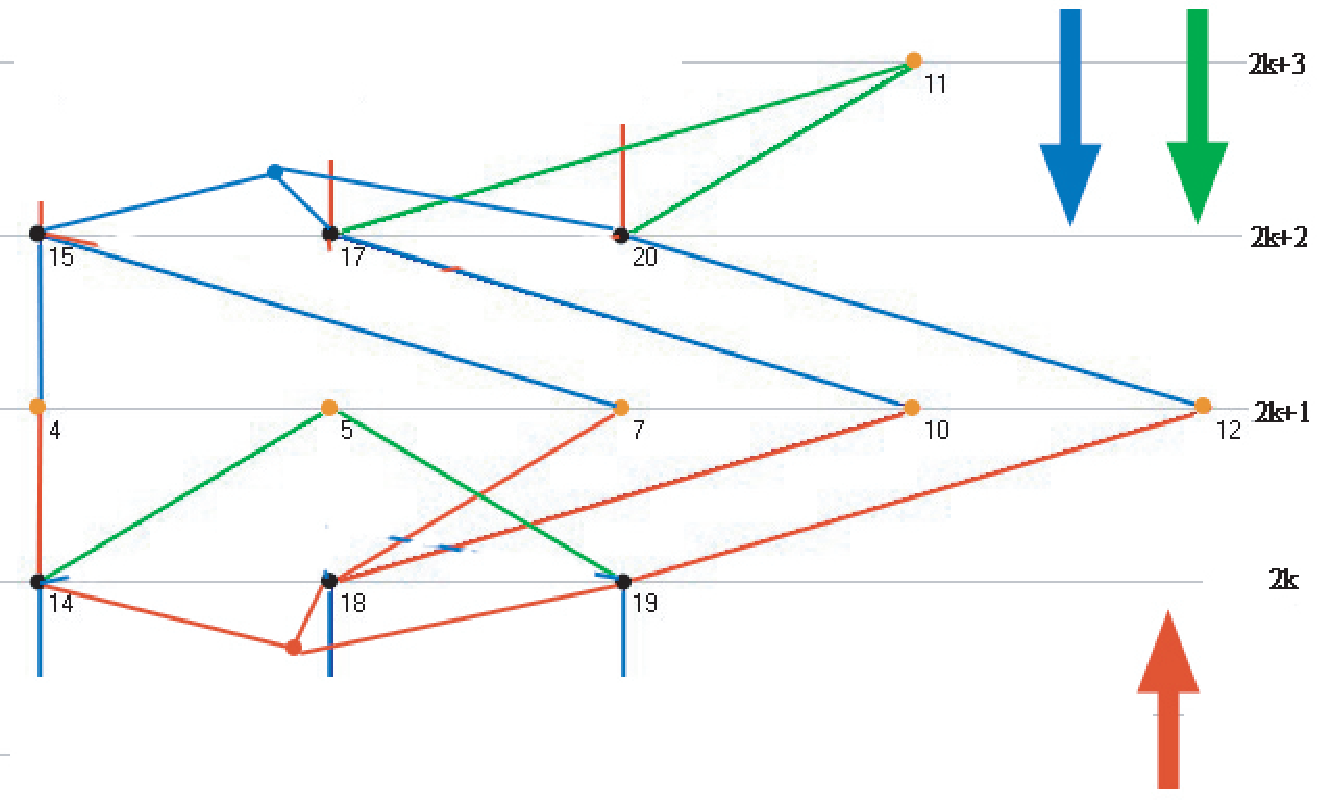}

\caption{$\gamma$-graph. \label{fig:F3-gamma-graph}}

\end{figure}

Let us consider subgraphs of $B$-graph, that subgraphs are defined
by any $\beta$-decomposition and formed from $\beta$-trees with
common orange vertices (Fig. \ref{fig:F3-gamma-graph}). In this subgraph
the black roots of red $\alpha$-trees are placed on the level $2k,\: k=0,1,2,...$,
the black roots of blue $\alpha$-trees are placed on the level $2k+2$,
the orange roots of green $\alpha$-trees are placed on the level
$2k+3$. Common vertices are leafs of blue and red trees. These vertices
are placed on the level $2k+1$. Such graphs are called $\gamma$-\textit{graphs.}

Clearly not more than two $\gamma$-graphs may correspond to every
even level $2k$: one of them may be formed from growing up red trees,
another may be formed from growing down blue trees. I.e. only one
$\gamma$-graph corresponds to zero level. This $\gamma$-graph is
formed from one red tree. Also only one $\gamma$-graph corresponds
to top level. This $\gamma$-graph is formed from blue and/or green
$\beta$-tree(s). Only one $\gamma$-graph corresponds to odd level
$2k+1$ always (roots of green trees of lower placed $\gamma$-graph
do not relate to given $\gamma$-graph). 
\begin{lem}
\textbf{\label{L3-gamma-graphs}} If in the result $D_{\beta}(t),\: t\in V$
for $B$-graph $G_{B}$ we obtain the set of $\gamma$-graphs $\{\Gamma_{1},\:\Gamma_{2},...\}$,
then $(\underset{(i)}{\bigcup}\Gamma_{i})\cap G_{B}=G_{B}$.\hfill{}$\square$\end{lem}
\begin{proof}
From Lemma \ref{L1-D-alfa} it follows that every edge and vertices
incident with this edge belong to any $\beta$-tree within $B$-graph.
Every $\beta$-tree belongs to any $\gamma$-graph by definition.
\end{proof}
\begin{flushleft}
Removing orange roots of green trees from $\gamma$-graph we obtain
$\delta$\textit{-graph} (Fig.\ref{fig:F4-delta-graph}). The root
of red $\beta$-tree and the root of blue $\beta$-tree within $\delta$\textit{-}graph
are called \textit{red and blue roots of $\delta$-graph}. If a root
of $\delta$\textit{-}graph is red or blue additional vertex, then
it is called \textit{additional root}. Also for following lemma we
note that $\gamma$-graphs and $\delta$-graphs are subgraphs and
we will use vertex codes (6) (see Definition \ref{D3-BNF}). These
codes include levels (3). So we will say about labeled $\delta$-graphs,
i.e. about $\delta$-\textit{graphs of the same level.} 
\begin{figure}[H]
\includegraphics[scale=0.4]{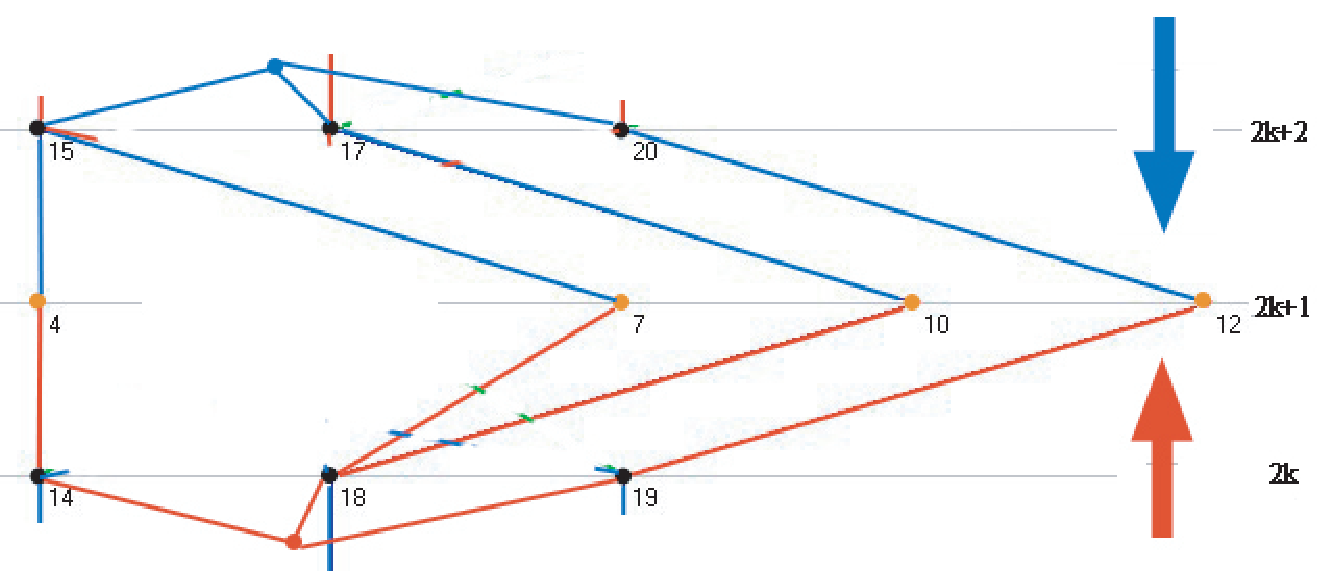}\caption{$\delta$-graph.\label{fig:F4-delta-graph}}

\end{figure}

\par\end{flushleft}
\begin{lem}
\textbf{\label{L8-delta-graphs-iso}} For isomorphism of the same
level $\delta$-graphs it is necessary and sufficient to have a one-to-one
correspondence between their vertex codes of red roots and blue roots
respectively.\hfill{}$\square$\end{lem}
\begin{proof}
Let graphs $G=(V,E)$ and $G^{\prime}=(V^{\prime},E^{\prime})$ be
$\delta$-graphs. Let $X=\{v_{1},v_{2},...,v_{k}\}$ and $X^{\prime}=\{v_{1}^{\prime},v_{2}^{\prime},...,v_{k}^{\prime}\}$
be the sets of orange vertices of these graphs respectively. Let $r_{b},\: r_{r},\, r_{b}^{\prime},\: r_{r}^{\prime}$
be blue and red roots respectively. Let $T_{b},\: T_{r},\, T_{b}^{\prime},\: T_{r}^{\prime}$
be blue and red trees respectively. 

Suppose that there is one-to-one correspondence between vertex codes
of blue and red roots respectively. Thus, tuples of $r_{b},\, r_{b}^{\prime}$
are equal, and tuples of $r_{r},\: r_{r}^{\prime}$ are equal. Hence,
from Theorem \ref{T2-tree-iso}, we see that $T_{b}\cong T_{b}^{\prime}$
and $T_{r}\cong T_{r}^{\prime}$. Hence, for every $v_{i}\in X$ we
can find similar vertex $v_{j}^{\prime}\in X$, and from Lemma \ref{L7-similar-tree-vert}
it follows that the tuples of these vertices and the tuples of their
ancestors (up to the root) are equal respectively. From Lemma \ref{L5-x-graphs}
it follows that $G\cong G^{\prime}$.

Now suppose that there is not one-to-one correspondence between vertex
codes of blue and/or red root(s). Hence, the tuples of $r_{b},\, r_{b}^{\prime}$
and/or the tuples of $r_{r},\: r_{r}^{\prime}$ are not equal. Hence,
red and/or blue trees are not isomorphic (Theorem \ref{T2-tree-iso}).
Clearly that if one of given graphs has a subgraph, which is not isomorphic
with every subgraph of other given graph, then given graphs are not
isomorphic. Hence, in this case, graphs $G$ and $G^{\prime}$ are
not isomorphic.
\end{proof}
Define the \textit{function}:
\[
f(M)=K,
\]

where $M=\{\Delta_{1},...,\Delta_{p}\}$ is a set of $\delta$-graphs
$\Delta_{i}$, $i=1,...,p$;

$K$ is multiset of tuple pairs $(t_{b}(i),\: t_{r}(i))$, $\mid K\mid=\mid M\mid=p$;

$t_{b}(i)$ is tuple of blue root of $\Delta_{i}$;

$t_{r}(i)$ is tuple of red root of $\Delta_{i}$.
\begin{lem}
\textbf{\label{L4-delta-graphs}} If in the result of all decompositions
$D_{\beta}(t_{i}),\: t_{i}\in V,\: i=1,2,...,n$ for $B$-graph $G_{B}$
we obtain the set of $\delta$-graphs $M=\{\Delta_{1},\:\Delta_{2},...\}$,
then $(\underset{(j)}{\bigcup}\Delta_{j})\cap G_{B}=G_{B}$, and $f(M)$
uniquely represents $G_{B}$.\hfill{}$\square$\end{lem}
\begin{proof}
For blue and red edges every $\delta$-graph looks like $\gamma$-graph
from which it was produced. Hence, we have the result from Lemma \ref{L3-gamma-graphs}.
Let us consider green edges. Let $D_{\beta}(h)$ produce green tree
with black leafs $i,j\in V$ on level $d$ and with orange root $k\in U$
on level $d+1$. So, $\delta$-graphs of level $d$ have not edges
$(i,k)$ and $(k,j)$. Suppose that all $\delta$-graphs for another
start vertices have not these edges also. Let us select vertex $i$
for start vertex. Then in the result of $D_{\beta}(i)$ vertex $i$
is placed on level 0, vertex $k$ is placed on level 1, vertex $j$
is placed on level 2, edge $(i,k)$ is red and edge $(k,j)$ is blue.
Respective $\delta$-graph has these edges. Hence our supposition
is not correct.

Let graph \foreignlanguage{russian}{$G_{B}$} isomorphic to graph
\foreignlanguage{russian}{$G_{B}^{\prime}$}. From Lemma \foreignlanguage{russian}{\ref{L2-beta-tree}},
we see that for each decomposition \foreignlanguage{russian}{$D_{\beta}(t_{i}),\: t_{i}\in V,\: i=1,2,...,n$}
we can find a decomposition \foreignlanguage{russian}{$D_{\beta}(t^{\prime}),\: t^{\prime}\in V^{\prime}$}
such that number of levels is the same and every $\beta$-tree of
\foreignlanguage{russian}{$G_{B}$} is isomorphic to $\beta$-tree
of \foreignlanguage{russian}{$G_{B}^{\prime}$} (for the same color
and the same level). Taking into account Corollary \ref{CC4} and
Lemma \foreignlanguage{russian}{\ref{L8-delta-graphs-iso}} we see
that \foreignlanguage{russian}{$f(M)=f(M^{\prime})$}, where \foreignlanguage{russian}{$M^{\prime}$}
is a set of all $\delta$-graphs for graph \foreignlanguage{russian}{$G_{B}^{\prime}$}.
\end{proof}
Removing additional roots from $\delta$-graph we obtain a graph (disconnected
in general case) consists blue and red trees. Such graph is called
$\sigma$-graph. Removing additional roots from a few $\delta$-graphs
we produce one disconnected $\sigma$-graph. Removing orange roots
of green trees within $G_{B}^{\prime}$ we obtain $\sigma$-graph
also.
\begin{lem}
\textbf{\label{L9-e-graphs-iso}} For isomorphism of $\sigma$-graphs
it is necessary and sufficient to have a one-to-one correspondence
between vertex codes of roots of red and blue trees respectively.\hfill{}$\square$\end{lem}
\begin{proof}
Clearly only one $\delta$-graph can be reconstructed from $\sigma$-graph
via adding blue and/or red root(s). If for two $\sigma$-graphs we
have a one-to-one correspondence between vertex codes of roots of
red and blue trees respectively, then we have the same correspondence
for reconstructed $\delta$-graphs. Hence from Lemma \ref{L8-delta-graphs-iso}
it follows that reconstructed $\delta$-graphs are isomorphic. If
$G\cong G'$ and vertices $v,\: v'$ are similar, then $G-v\cong G'-v'$
(Corollary \ref{=0004211-similar-vert}), thus removing blue and/or
red roots from isomorphic $\delta$-graphs produces isomorphic $\sigma$-graphs.\end{proof}
\begin{lem}
\begin{flushleft}
For isomorphism of graphs $G_{B}$ and $G_{B}^{\prime}$ it is necessary
and sufficient to have $S=S^{\prime}$\textup{ (Algorithm \ref{alg:A1})}.\hfill{}$\square$
\par\end{flushleft}\end{lem}
\begin{proof}
From every $\sigma$-graph may be reconstructed only one $\delta$-graph.
From Lemma \ref{L4-delta-graphs} it follows that sets of these $\delta$-graphs
uniquely represents $G_{B}$ and $G_{B}^{\prime}$ respectively. So,
let us consider corresponded $\sigma$-graphs. 

In $i$-th step of loop 2 (Algorithm \ref{alg:A1}) for start vertex
$i$ of graph $G_{B}$ we obtain $\sigma$-graph $\sigma_{i}$. If
$G\cong G'$, then analogous graph $\sigma_{j}^{\prime}$ is produced
in $j$-th step of of loop 2 for graph $G_{B}^{\prime}$. From Lemma
\ref{L9-e-graphs-iso} it follows that for isomorphism of graphs $\sigma_{i}$
and $\sigma_{j}^{\prime}$ it is necessary and sufficient to have
a one-to-one correspondence between vertex codes of roots of red and
blue trees respectively.

For all start vertices $v_{i}\in V,\: i=1,...,n$ loop 2 produces
all possible $\sigma$-graphs. After sorting vertex codes and writing
them to vectors $S$ and $S^{\prime}$ respectively we have following
two cases. If $S=S^{\prime}$, then for every $\sigma_{i}$ we can
find $\sigma_{j}^{\prime}$ with corresponded vertex codes, i.e. isomorphic.
Otherwise, if $S\neq S^{\prime}$, then for some $\sigma_{i}$ we
can not find $\sigma_{j}^{\prime}$ with corresponded vertex codes. 
\end{proof}
\begin{flushleft}
The defect of Algorithm \ref{alg:A1} is too long strings. Indeed,
for example, in the case of regular graph with vertex degree $k$
and diameter $d$ the length $l$ of string $c(v_{j})$ for vertex
code in step 12 may be estimated as
\par\end{flushleft}

\begin{flushleft}
\[
l\approx(2ka)^{d},
\]

\par\end{flushleft}

where $a$ is length of string represents simple vertex code (because
we speak about approximate estimation we select a vertex with maximal
$a$). 

\begin{flushleft}
If we imagine that all simple vertex codes have the same length $a$,
then, neglecting terminal symbols, whose contribution is not very
important, we see that every iteration of loop 7 multiplies edge code
length by 2 and vertex code length by $k$ times. To overcome this
problem we use simple trick: add common vertex to source graph and
link other vertices with this common vertex. Clearly the diameter
of produced graph is not more than 2. From Corollary \ref{D1-similar-vert}
it follows that if such graphs are isomorphic, then removing of common
vertices produces isomorphic graphs also. Now we introduce main algorithm
(Algorithm \ref{alg:A2}). 
\begin{algorithm}[H]
\begin{enumerate}
\item if number of vertices of graph $G_{s}$ and number of vertices of
graph $G_{s}^{\prime}$ are different, then graphs are not isomorphic,
exit; 
\item if number of edges of graph $G_{s}$ and number of edges of graph
$G_{s}^{\prime}$ are different, then graphs are not isomorphic, exit; 
\item add a common vertex to graph $G_{s}$; 
\item add a common vertex to graph $G_{s}^{\prime}$;
\item call procedure $P_{C}$ to calculate vector $S$ for graph $G_{s}$;
\item call procedure $P_{C}$ to calculate vector $S^{\prime}$ for graph
$G_{s}^{\prime}$;
\item if $S=S^{\prime}$, then graphs are isomorphic, else graphs are not
isomorphic. 
\end{enumerate}
\caption{\label{alg:A2}Graphs $G_{s}$ and $G_{s}^{\prime}$ isomorphism testing.}
\end{algorithm}

\par\end{flushleft}

To estimate computation complexity of Algorithm \ref{alg:A2} for
the worst case we have to consider the most hard procedure $P_{C}$
(Algorithm \ref{alg:A1}). From Theorem \ref{T1-tuple-time} it follows
that $\mathrm{rgb}$ label calculation (step 5) requires linear time
proportional to $n$. The step 6 has the same dependence. Statements
9,10 are the most hard. Loops 2 and 7 repeat these statements not
more than $2n$ (i.e. $\mathrm{dm}(G_{s})\leqslant2)$. Statement
9 is a loop repeated $m$ times. Statement 10 is a loop repeated $n$
times. However, there is comparison of only two vertex codes in statement
9. In contrast, there is sorting up to $n$ vertex codes in statement
10. There are many effective algorithms of sorting require less than
$n^{2}$ comparisons of sorted elements in literature. So, statement
9 requires $2m$ comparisons and statement 10 requires not more than
$n^{3}$ comparisons. Taking into account that maximal number of edges
within a graph is number of edges of complete graph, i.e. $(n^{2}-n)/2$,
we see that statement 10 is the most hard. Taking into account loops
2 and 7, we see that total number of comparisons is not more than
$2n^{5}$. Multiplying this value by the length of string $l$, we
obtain total number of symbolic comparisons $p$ (Remark \ref{N1-cmp-array}):
\[
p=2n^{5}l\approx2n^{5}(2ka)^{2}.
\]

Assume $k\leqslant n$ and $a<cn$, where $c$ is constant equals
number of bits necessary for representation of one symbol of a string.
Thus

\[
p<8cn^{9}.
\]

Statements 14,15,16 require less number of comparisons. Hence, neglecting
the lowest terms and factors, the total complexity of the algorithm
can be estimated as $O(n^{9})$.

\section*{Conclusion}

The essence of this work is a method of reduction of general task
of graph isomorphism testing to more particular task of labeled trees
isomorphism, that task was solved earlier. Perhaps, some of used definitions,
algorithms and proofs look like a little redundant. And perhaps, introduced
data structures have too large size. However, the main goal of this
work is theoretic result, for that redundancy is better than insufficiency.
Introduced algorithm answers (''yes'' or ''no'') question about
isomorphism of pair of given graphs, but in the case of positive answer
the algorithm does not produce any possible isomorphism in output.
The algorithm was implemented in Borland Delphi-7 for MS Windows.
The source code and the executables are available via 

http://mt2.comtv.ru/

The password to unzip is

hH758-kT402-N3D8a-961fQ-WJL24

Also translation into Russian is available via this URL.

Some $B$-graph properties were not used for proving, but these properties
may be useful for this approach progress. So, the properties are described
in Appendix 1.

\section*{Acknowledgments}

Many thanks to Gennadiy M. Hitrov (Saint Petersburg University) for
discussion of this paper. Many thanks to Mary P. Trofimov for help
in preparation of the manuscript.

\section*{Appendix 1}

The adjacency matrix of graph $G_{B}$ has the form:

\[
A=\left(\begin{array}{cc}
O & B\\
B^{T} & O
\end{array}\right),
\]

where $O$ is an all-zero matrix;

$B$ is $m\times n$ (0,1)-matrix:

\[
B=\left\Vert b_{ij}\right\Vert ,
\]

where $b_{11}$is an adjacency for the first vertex from $U$ (i.e.
additional vertex 1) with the first vertex from $V$ (i.e. vertex
$m+1$) etc.

Let us exclude trivial cases $n\leqslant3$ from following discussion.
Also we will consider connected graphs only.
\begin{prop}
\label{A1}If we interchange two rows (columns) of matrix $B$, then
this interchange preserves adjacency.\hfill{}$\square$\end{prop}
\begin{proof}
Let us interchange rows $i$, $j$. In the result we have $b_{ik}=b_{jk}^{\prime}$
and $b_{jk}=b_{ik}^{\prime}$, where $b_{jk}^{\prime}$ and $b_{ik}^{\prime}$
are matrix elements before the interchange; $b_{jk}$ and $b_{ik}$
are matrix elements after the interchange. This means that if the
graph initially had an edge $(i,k),\: i\in U,\: k\in V$, then this
edge is denoted by $(j,k),\: j\in U$ after the interchange. And if
the graph initially had an edge $(j,k),\: j\in U,\: k\in V$, then
this edge is denoted by $(i,k),\: i\in U$ after the interchange.
The same situation is observed for all edges which are incident with
vertices $i$, $j$ respectively. Hence, in the result only vertices
numbers $i$, $j$ are interchanged, but the adjacency is preserved.

Similarly let us interchange columns $i$, $j$. In the result we
have $b_{ki}=b_{kj}^{\prime}$ and $b_{kj}=b_{ki}^{\prime}$. This
means that if the graph initially had an edge $(i,k),\: i\in V,\: k\in U$,
then this edge is denoted by $(j,k),\: j\in V$ after the interchange.
And if the graph initially had an edge $(j,k),\: j\in V,\: k\in U$,
then this edge is denoted by $(i,k),\: i\in V$ after the interchange.
The same situation is observed for all edges which are incident with
vertices $i$, $j$ respectively. Hence, in the result only vertices
numbers $i$, $j$ are interchanged, but the adjacency is preserved.\end{proof}
\begin{prop}
\label{A2}No interchanges of rows and columns of matrix $B$ can
produce following block:

\begin{equation}
\left(\begin{array}{cc}
1 & 1\\
1 & 1
\end{array}\right).\label{eq:11}
\end{equation}
\hfill{}$\square$\end{prop}
\begin{proof}
Suppose that such block is possible for vertices $i,j\in V$ and $p,q\in U$.
Hence, there are two edges $(i,j)$ in source graph. But it is not
possible by condition (we do not consider multigraphs). This contradiction
shows that our supposition is not correct.\end{proof}
\begin{prop}
\label{A3}Equal rows or equal columns are impossible for matrix $B$.
\hfill{}$\square$\end{prop}
\begin{proof}
Suppose that two rows of matrix $B$ are equal. Every additional vertex
has degree 2. Thus every row has exactly two units. If any rows are
equal, then we can produce block (\ref{eq:11}) via interchanges of
rows and columns, that contradicts Proposition \ref{A2}. Hence our
supposition is not correct.

Now let us consider following cases for columns. 

1) Two equal columns are all-zero columns. This is not possible, because
we consider only connected graphs by condition.

2) Columns $i$, $j$ are equal and there is only one unit in every
of these columns. In this case, we have row $k$  such that $b_{ki}=b_{kj}=1$.
Since we do not consider trivial cases ($n\leqslant3$), we have graph
that has subgraph consisted only one edge $(i,j)$, where vertices
$i$, $j$ are disconnected with other vertices. But it contradicts
the condition that only connected graphs have to be considered.

3) Columns are equal and every of these columns have not less than
two units. In this case, we can produce block (\ref{eq:11}) via interchanges
of rows and columns that contradicts Proposition \ref{A2}. Hence
our supposition is not correct.
\end{proof}
From Proposition \ref{A1} follows supposition that if we sort matrix
$B$ by rows, by columns, and again by rows and by columns etc., until
matrix stops change, then we obtain ''maximal matrix'' independent
on vertex numbers. Unfortunately, simple counter-examples show that
this supposition is not correct.

\bibliographystyle{plain}
\bibliography{Ref-en}

\end{document}